\documentclass[reprint,prb,aps,superscriptaddress,showpacs,colors]{revtex4-1}
\usepackage{epsfig,color}
\usepackage{graphicx}
\usepackage{amsmath}

\begin{document}
\title{Spin polarization anisotropy\\ in a narrow spin-orbit-coupled nanowire quantum dot}

\author{M. P. Nowak}
\affiliation{AGH University of Science and Technology, Faculty of Physics and Applied Computer Science,\\
al. Mickiewicza 30, 30-059 Krak\'ow, Poland}
\author{B. Szafran}
\affiliation{AGH University of Science and Technology, Faculty of Physics and Applied Computer Science,\\
al. Mickiewicza 30, 30-059 Krak\'ow, Poland}

\date{\today}


\begin{abstract}
One and two-electron systems confined in a single and coupled quantum dots defined within a nanowire with a finite radius are studied in the context of spin-orbit coupling effects. Anisotropy of the spin-orbit interaction is discussed in terms of the system geometry and orientation of the external magnetic field vector. We find that there are easy and hard spin polarization axes and in the quantum dot with strong lateral confinement electron spin becomes well defined in spite of the presence of spin-orbit coupling. We present an analytical solution for the one-dimensional limit and study its validity for nanowires of finite radii by comparing the results with a full three-dimensional calculation. The results are also confronted with the recent measurements of the effective Land\'e factor and avoided crossing width anisotropy in InSb nanowire quantum dots [S. Nadj-Perge {\it et al.}, Phys. Rev. Lett. 108, 166801 (2012)].

\end{abstract}
\maketitle

\section{Introduction}
There is a growing interest in gated semiconductor nanowires in the context of possible applications for spin-operating devices.\cite{nadj-perge,schroer,nadj2012,frolov} These structures provide a good basis for creation of small electrostatic quantum dots with confinement introduced by external potentials. Energy spectra of such dots as determined\cite{sostrnanowire} by transport spectroscopy bear distinct signatures of strong spin-orbit (SO) interaction which results from the structure inversion asymmetry (Rashba SO coupling\cite{rashba}) or the bulk inversion asymmetry (Dresselhaus SO interaction\cite{dresselhaus}). SO coupling mixes spin and orbital degrees of freedom thus opening the possibility of fully electrical control of the electron spin.\cite{nadj-perge,schroer,nadj2012,edsr,frolov,sfet} Moreover SO coupling allows for electron spin relaxation mediated by phonons,\cite{phonons,phonons2} and introduces anisotropic corrections to spin exchange interaction for electrons in double quantum dots.\cite{spinswap}

The SO coupling opens avoided crossings\cite{sostrnanowire} in the quantum dot energy spectra as a function of the external magnetic field ($\mathbf{B}$).
The width of the avoided crossings between energy levels of different spin states depends on the orientation of $\mathbf{B}$ vector,
which reveals the spatial anisotropy of the SO interaction.\cite{nadj2012,japac,nowak2011,nowak2011c}
Moreover, the mixing of the spin states by SO coupling determines an effective Land\'e factor ($g$-factor) and its anisotropy\cite{japgfac} as a function of the magnetic field orientation.


It is well known, that in the presence of SO coupling the electron spin can be well defined in the stationary eigenstates
only for equal Rashba and Dresselhaus SO coupling constants.\cite{nsfet}
This fact was exploited in a proposal of nonballistic spin field effect transistor\cite{nsfet} and for prediction \cite{bernevig} of persistent spin helix.\cite{koralek}
In the present work we demonstrate that in the limit of strong lateral confinement the electron spins confined in the quantum dot become well defined in the direction
perpendicular to the wire axis and the external electric field vector in spite of the presence of the Rashba coupling.
We show that in a general case, the extent of the electron spin polarization strongly depends on the orientation of {\bf B} reflecting the anisotropy of SO interaction.

For a description of narrow nanowires a one-dimensional model is commonly used.\cite{1dused} In this work we present an analytical form of eigenstates for this approximation for a quantum dot defined in a nanowire. The analytical form of the SO-coupled wavefunctions accounts for the anisotropic spin polarization and explains a different strengths of the spin-splittings for varied orientation of the magnetic field. We study applicability of the one-dimensional model for a nanowire with a finite radius by comparing its results with the three-dimensional calculation for various geometries of the nanowire quantum dot. In order to relate the model results to the experimental measurements we study coupled two-electron quantum dots, i.e.  the configuration that is used for EDSR and the spin exchange experiments. The obtained shape of the $g$-factor and avoided crossing width dependence on magnetic field orientation resembles the findings of the experiment of Ref. \onlinecite{nadj2012} on InSb nanowire quantum dots.

\section{Theory}
\begin{figure}[ht!]
\epsfxsize=87mm
                \epsfbox[15 590 579 807] {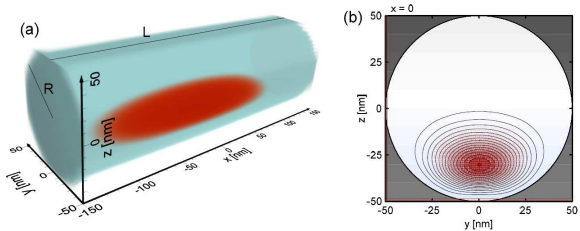}
                 \caption{(color online) (a) Sketch of the confinement potential $V(\textbf{r})$ of the nanowire quantum dot (with blue) and the single-electron charge density (with red) calculated for $F_z=10$ kV/cm. (b) Cross section of the confinement potential and the charge density for $x=0$.}
\label{pot}
\end{figure}

We consider a single-electron quantum dot defined in a narrow nanowire described by the three-dimensional Hamiltonian,
\begin{equation}
h = \frac{\hbar^2\textbf{k}^2}{2m^*} + V(\textbf{r}) + H_{SO} + \frac{1}{2}g\mu_B \mathbf{B}\cdot\sigma,
\label{3dham}
\end{equation}
where $\mathbf{k}=-i\nabla+e\mathbf{A}/\hbar$ with the gauge $\mathbf{A} = B (z \sin\phi,0,y \cos\phi)$. The magnetic field is aligned in the $xy$-plane with an angle $\phi$ between the $\mathbf{B}$ and $x$-axis -- in such a case the Zeeman term stands $\frac{1}{2}g\mu_B \mathbf{B}\cdot\sigma = \frac{1}{2}\mu_B g B(\sigma_x \cos\phi + \sigma_y \sin \phi)$,
$V(\textbf{r})$ stands for the confinement potential which we take in a separable form $V(\mathbf{r})=V_l(y,z)+V_L(x) + |e| \mathbf{F} \cdot \mathbf{r}$ where $V_l(y,z)$ is a 400 meV deep two-dimensional circular quantum well of radius $R$, $V_L(x)$ is a infinite quantum well with width $L$ (see Fig. \ref{pot}) and  $\mathbf{F} $ stands for the external electric field.
We account for Rashba SO coupling $H_{SO} = \alpha_0 \frac{\partial V}{\partial r} \cdot (\sigma \times \textbf{k})$ as the main SO interaction type in the [111] grown InSb nanowires.\cite{nadj2012} Unless stated otherwise we assume the electric field $\mathbf{F}=(0,0,F_z)$ with non-zero component in the $z$-direction (perpendicular to the axis of the wire) due to the gating of the nanowire.\cite{nadj-perge,schroer,nadj2012,frolov}
We assumed a hard-wall confinement potential of the wire. The electron wavefunction vanishes at the edge of circular quantum well $V_l(y,z)$ [see Fig. \ref{pot}(b)]. Therefore, the only part of the potential whose
gradient overlaps with the wave function and thus gives rise to the SO coupling effect is the external electric potential, i.e.  $H_{SO}=\alpha(\sigma_x k_y - \sigma_y k_x)$ where $\alpha=\alpha_0 F_z$.

To solve the Schr\"odinger equation we rewrite the Hamiltonian Eq.(\ref{3dham}) as $h=h_x+h_y+h_z+h_{ns}$, where

\begin{equation}
h_x=-\frac{\hbar^2}{2m^*}\frac{\partial^2}{\partial x^2}+V_L(x),\label{hs1}
\end{equation}

\begin{equation}
h_y=-\frac{\hbar^2}{2m^*}\frac{\partial^2}{\partial y^2} + V_B(y) + \frac{e^2B^2}{2m^*}y^2\cos^2\phi,\label{hs2}
\end{equation}

\begin{equation}
h_z=-\frac{\hbar^2}{2m^*}\frac{\partial^2}{\partial z^2} + V_B(z) + \frac{e^2B^2}{2m^*}z^2\sin^2\phi + |e|F_zz,\label{hs3}
\end{equation}
are separable in the $x,y$ and $z$ directions spin-independent parts. The infinite quantum wells $V_B(y)$ and $V_B(z)$ of width $2R$ define the computational box
 and
\begin{equation}
\begin{split}
h_{ns}&=-\frac{i\hbar eB}{m^*}\left(z\sin\phi\frac{\partial}{\partial x}+y\cos\phi\frac{\partial}{\partial z}\right) \\
&+\frac{1}{2}g\mu_bB\left[ \sigma_x \cos\phi + \sigma_y\sin\phi \right] + H_{SO} + V_l(y,z),
\end{split}
\end{equation}
is the nonseparable part that contains the spin dependency and the potential of the cylindrical quantum well $V_l(y,z)$.

The calculation procedure proceeds as follows. We calculate eigenvectors of $h_x, h_y$ and $h_z$ on meshes containing 1000 points and use them for construction of a basis (which consist of $N=8192$ elements) in which $h$ Hamiltonian is further diagonalized. As a result we obtain three-dimensional spin-orbitals $\psi(\mathbf{r},\sigma)$.
Note, that introducing the infinite quantum wells $V_B$ in the first step fixes the basis for the diagonalization of the complete Hamiltonian.

The solutions of two-electron system described by the Hamiltonian,
\begin{equation}
H=h_1+h_2+\frac{e^2}{4\pi\varepsilon_0\varepsilon |\textbf{r}_1-\textbf{r}_2|},
\label{2eham}
\end{equation}
are found in the basis constructed from products of antisymmetrized single-electron spin-orbitals $\psi(\mathbf{r},\sigma)$,
\begin{equation}
\begin{split}
\Psi(\mathbf{r}_1,\sigma_1,&\mathbf{r}_2,\sigma_2)=\\
&\frac{1}{\sqrt{2}}\sum_{i=1}^N\sum_{j=i+1}^Nc_{ij}[\psi_i(\mathbf{r}_1,\sigma_1)\psi_j(\mathbf{r}_2,\sigma_2)\\
&-\psi_i(\mathbf{r}_2,\sigma_2)\psi_j(\mathbf{r}_1,\sigma_1)],
\end{split}
\end{equation}
where the coefficients $c_{ij}$ are found by diagonalization of Hamiltonian Eq. (\ref{2eham}) according to the configuration interaction method with $N=20$. The scheme treats the Coulomb interaction in an exact manner. For the calculation of the Coulomb matrix elements we use two-step method that replaces six-dimensional integrations
by calculation of the Poisson equation 
for the potential generated from single-electron wavefunctions and integrate it with the product of wavefunction of the other electron.\cite{nowak2011}

We adopt material parameters\cite{params} for InSb, namely : $m^*=0.014m_0$, $g=-51$, $\varepsilon = 16.5$, and $\alpha_0=5\;\mathrm{nm}^2$. In the bulk of the paper we choose $F_z = 50$ kV/cm which results in SO interaction constant $\alpha=25$ meVnm. Unless stated otherwise we take $L=300$ nm.

\section{Results}
\subsection{Single electron in a finite thickness nanowire quantum dot}
\begin{figure}[ht!]
\epsfxsize=75mm
                \epsfbox[26 193 555 666] {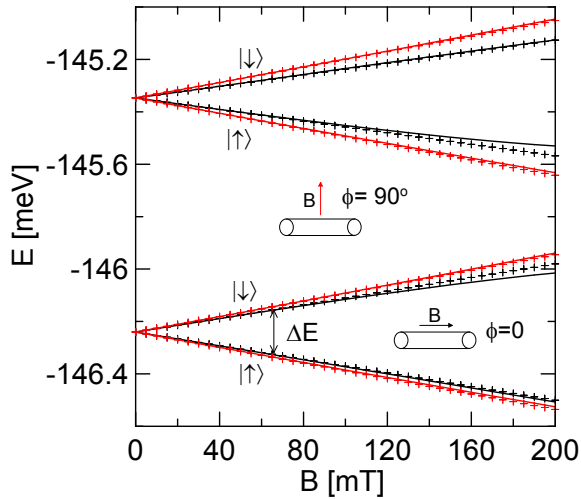}
                 \caption{(color online) Single-electron energy spectrum for the SO coupled nanowire quantum dot with radius $R=50$ nm and SO interaction constant $\alpha=25$ meVnm plotted with lines for two orientations of the magnetic field. The crosses are the results obtained from asymptotic one-dimensional solution -- see text. With $|\uparrow\rangle$ and $|\downarrow\rangle$ we mark the spin-polarization of the states parallel and antiparallel to the magnetic field respectively as found without SO coupling.}
\label{fig0}
\end{figure}

\begin{figure}[ht!]
\epsfxsize=75mm
                \epsfbox[40 28 573 821] {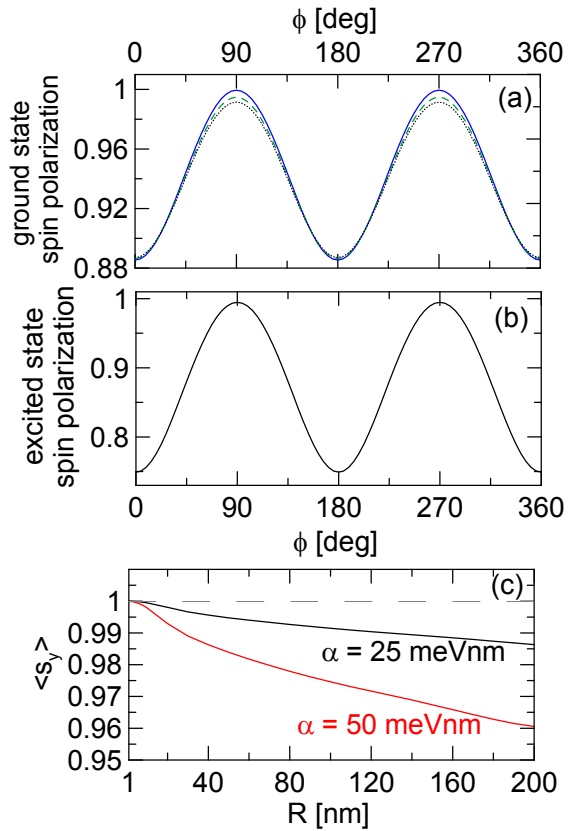}
                 \caption{(color online) (a) Mean value of the spin along the magnetic field direction obtained for the ground state of the nanowire quantum dot with radius $R=10$ nm (blue solid curve), $R=50$ nm (green dashed curve) and $R=100$ nm (black dotted curve). (b) Spin polarization of the second excited state for $R=50$ nm. (c) Mean value of the spin-$y$ component for the magnetic field aligned along the $y$-direction as a function of the nanowire radius $R$. (a), (b) and (c) are obtained for $B=5$ mT. Results for $\alpha=50$ meVnm are obtained with $F_z=100$ kV/cm.}
\label{fig1}
\end{figure}

Lowest part of the energy spectrum of the single-electron quantum dot is presented in Fig. \ref{fig0}. In the absence of the magnetic field all the levels are Kramer's doublets. We include residual magnetic field $B=5$ mT and inspect the spin polarization along the magnetic field direction (calculated as $\langle s_B \rangle = \langle s_x \rangle \cos(\phi) + \langle s_y \rangle \sin(\phi)$). In Fig. \ref{fig1}(a) we observe that the spin polarization undergoes oscillatory changes as a function of $\mathbf{B}$ orientation. This reflects the presence of an easy and hard spin polarization axes in the system. For the magnetic field oriented perpendicular to the nanowire axis the spin is easily polarized -- taking values close to 1 $[\hbar/2]$. On the other hand for $\mathbf{B}$ oriented along the wire the $\langle s_B \rangle$ is around 0.885 $[\hbar/2]$. The amplitude of the oscillations depends on the nanowire radius (compare the curves in Fig. \ref{fig1}(a) for three values of R) and the oscillations are the strongest for narrow nanowire with $R=10$ nm. The spin polarization of the excited state is presented in Fig. \ref{fig1}(b). We observe that the amplitude of the oscillation is stronger than the one obtained for the ground state but the spin polarization for $\phi=90^o$ is again close to 1 $[\hbar/2]$.

Let us inspect the degree of the maximal spin polarization at the easy axis  $\phi=90^o$. In Fig. \ref{fig1}(c) we plot mean value of the spin-$y$ component of the ground state versus the wire radius R. We observe that as the wire becomes narrower the spin polarization becomes almost complete (i.e., $1-\langle s_y \rangle 2/\hbar < 10^{-4}$ for $R=1$ nm) despite the presence of the SO coupling. Existence of directions in which the spin can be exactly polarized should facilitate the qubit initialization and increase the spin coherence times.
On the other hand as the wire becomes wider the spin-polarization drops with the slope of the curves in Fig. \ref{fig1}(c) depending on the SO coupling constant $\alpha$. Note that the extent of the wavefunction in the $z$-direction is limited also by the applied electric field.

When the magnetic field is increased it splits the doublets -- see the energy levels in Fig. \ref{fig0}. The energy splittings obtained for the magnetic field perpendicular to the nanowire axis (red curves in Fig. \ref{fig0}) are stronger than the ones obtained for the magnetic field parallel to the nanowire axis (black curves in Fig. \ref{fig0}). In the following we explain this observation.

\subsection{Asymptotic solution (1D limit)}
When the wire becomes narrow the energy of the states excited in the radial direction rises.
 It is reasonable then to inspect the case where the radial degrees of freedom are decoupled from the longitudinal one (the $x$-direction). Such a system is described by the one-dimensional Hamiltonian,\cite{1dused,nowak2012}

\begin{equation}
h_{1D} = \frac{\hbar^2 k_x^2}{2m^*} + V_L(x) - \alpha \sigma_y k_x + \frac{1}{2}\mu_B g B(\sigma_x \cos\phi + \sigma_y \sin \phi),
\label{1dham}
\end{equation}
where $k_x = -i \frac{\partial}{\partial x}$.

Generally, the analytical solution for a SO coupled confined systems are not known with the exception of a special case of equal strength of Rashba and Dresselhaus coupling described in Ref. \onlinecite{nsfet}. Here we note however that in the absence of the magnetic field ($B=0$) the Hamiltonian (\ref{1dham}) commutes with spin-$y$ Pauli matrix and its eigenstates are the states with well defined spin in the $y$-direction. We find that for a quasi one-dimensional nanowire the spin-orbitals (where $N$ stands for the orbital quantum number and $\pm$ for the spin polarization of the state) have the form,
\begin{equation}
\Psi_{N\pm} = \frac{1}{\sqrt{2}} \left( \begin{array}{c} 1\\ \pm i \end{array} \right) \varphi_N(x) \exp\left[ \pm \frac{i\alpha m^*}{\hbar^2} x \right],
\label{soieigenspinors}
\end{equation}
where $\varphi_N(x)$ are spin-independent eigenstates of Hamiltonian (\ref{1dham}) for $\alpha=0$ and $B=0$. The eigenenergies of the Hamiltonian (\ref{1dham}) are $E_{1D}=E_{\alpha=0,N}+E_{SO}$ where $E_{SO}=-\alpha^2m^* / (2\hbar^2)$ is the energy shift to the whole energy spectrum introduced by the SO interaction\cite{enote} and $E_{\alpha=0,N}$ is an energy level of the $N$'th eigenstate obtained without SO coupling.

The magnetic field affects the energy levels of a strongly confined electron mainly through the Zeeman spin-splitting. To investigate its influence on the SO eigenstates with an orbital excitation $N$ let us diagonalize $h_{1D}$ for $B>0$ in a basis consisting of a degenerate pair $\Psi_{N+}$ and $\Psi_{N-}$. The Hamiltonian matrix is
\begin{equation}
\left( \begin{array}{cc}
\langle \Psi_{N+} | h_{1D} | \Psi_{N+} \rangle & \langle \Psi_{N-} | h_{1D} | \Psi_{N+} \rangle \\
\langle \Psi_{N+} | h_{1D} | \Psi_{N-} \rangle & \langle \Psi_{N-} | h_{1D} | \Psi_{N-} \rangle \end{array} \right),
\label{matrix}
\end{equation}
where the diagonal elements are defined as follows
\begin{equation}
\langle \Psi_{N\pm} | h_{1D} | \Psi_{N\pm} \rangle = E_{1D} \pm \frac{1}{2} g \mu_B B \sin \phi,
\end{equation}
while the off-diagonal elements are
\begin{equation}
\begin{split}
&\langle \Psi_{N\pm} |  h_{1D} | \Psi_{N\mp} \rangle = \\& \mp i \frac{1}{2} g \mu_B B \int |\varphi_N|^2 \left[ \cos(\frac{2\alpha m^*}{\hbar^2}x) \mp  i\sin(\frac{2\alpha m^*}{\hbar^2}x) \right] dx \cos \phi.
\end{split}
\end{equation}
Let us denote $\lambda_N \equiv \int |\varphi_N|^2 \cos(\frac{2\alpha m^*}{\hbar^2}x) dx$ and $\kappa_N \equiv i\int |\varphi_N|^2 \sin(\frac{2\alpha m^*}{\hbar^2}x) dx$.

\begin{figure}[ht!]
\epsfxsize=75mm
                \epsfbox[28 108 566 740] {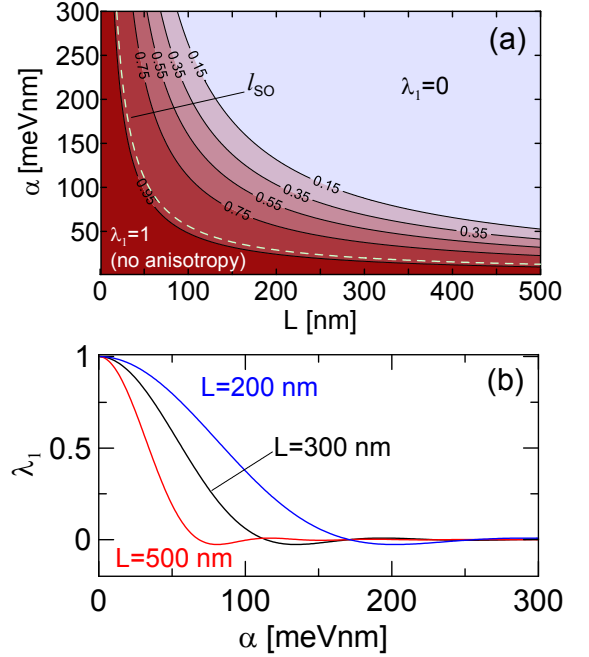}
                 \caption{(color online) Parameter $\lambda_1$ as a function of the dot length $L$ and SO coupling constant $\alpha$. (b) Cross-section of (a) for three different dot lengths $L$.}
 \label{fig2}
\end{figure}

The eigenstates of the matrix (\ref{matrix}) are
\begin{equation}
E_{N\pm} = E_{1D} \pm \frac{1}{2}g\mu_B B \sqrt{1-(1-\lambda_N^2+\kappa_N^2)\cos^2\phi}.
\label{anel}
\end{equation}
The energy difference between the states depends on the orientation of the magnetic field (angle $\phi$) as well as the parameters $\lambda_N$ and $\kappa_N$ that control the strength of the anisotropy of the spin splittings for rotated magnetic field. For the symmetric infinite quantum well confinement along the wire ($x$-direction) we obtain,\cite{comment1}
\begin{equation}
\lambda_1=\frac{\hbar^6\pi^2\sin(L\alpha m^*/\hbar^2)}{\alpha m^* L (\pi^2 \hbar^4-\alpha^2{m^*}^2L^2)},
\end{equation}
and
\begin{equation}
\lambda_2=\frac{4 \hbar^6 \pi^2\sin(L\alpha m^*/\hbar^2)}{\alpha m^* L (4\pi^2\hbar^4-\alpha^2{m^*}^2L^2)},
\end{equation}
and $\kappa_1=\kappa_2=0$ for the two lowest orbital states. The $\lambda_N$ depends on the quantum dot length and the SO strength. In Fig. \ref{fig2} we present the $\lambda_1$ parameter as a function of $L$ and $\alpha$. With the light-green dashed curve we depict the SO length $l_{SO}=\hbar/(m^*\alpha)$. We observe that $\lambda_1$ drops quickly when the length of the dot becomes greater than SO length. The shape of the $\lambda_1$ dependence on the SO strength for different quantum dot lengths is presented in Fig. \ref{fig2}(b) showing that the SO effects strongly depends on the quantum dot geometry and that $\lambda_1$ goes to 1 for vanishing SO coupling

The smaller $\lambda_N$ the stronger the SO coupling effects are. In particular for the magnetic field parallel to the nanowire axis the energy of the spin splitting is $E_S=g\mu_BB\lambda_N$. Consequently the splitting can even go to 0 due to strong mixing of the spin states by the SO interaction [the light blue region in Fig. \ref{fig2}(a)].

When the magnetic field is aligned in the direction perpendicular to the nanowire axis, i.e. $\phi=90^o$ or $\phi=270^o$ the off-diagonal elements of the matrix (\ref{matrix}) vanish and the energy levels are split by Zeeman energy with the bulk value of the $g$-factor. This is the reason for stronger spin splittings of the red curves in Fig. \ref{fig0}. For this configuration the spin-orbitals are separable into spin and orbital parts despite the presence of SO interaction and they have the exact form of Eq. (\ref{soieigenspinors}). For any other orientation of the magnetic field the off-diagonal elements mix the eigenstates (\ref{soieigenspinors}). This results in decreasing the spin splittings by the SO interaction by an amount that depends on  $\lambda_N$ and $\kappa_N$ parameters -- the spatial extent of the wave function along the nanowire and the strength of the SO coupling. Moreover the electron spin is no longer well defined as the electrons spin and orbital degrees of freedom are entangled.

We plot the energy spectrum obtained from Eq. (\ref{anel}) (shifted to match the energies obtained in the three-dimensional calculation at $B=0$) with the crosses in Fig. \ref{fig0}. The spin splitting obtained from the one-dimensional model well describes the results of the three-dimensional calculation. The only discrepancy is visible for the energy levels of the first and the second excited states for $B>100$ mT which is due to mixing of this two states by the SO interaction.

\begin{figure}[ht!]
\epsfxsize=75mm
                \epsfbox[34 170 560 690] {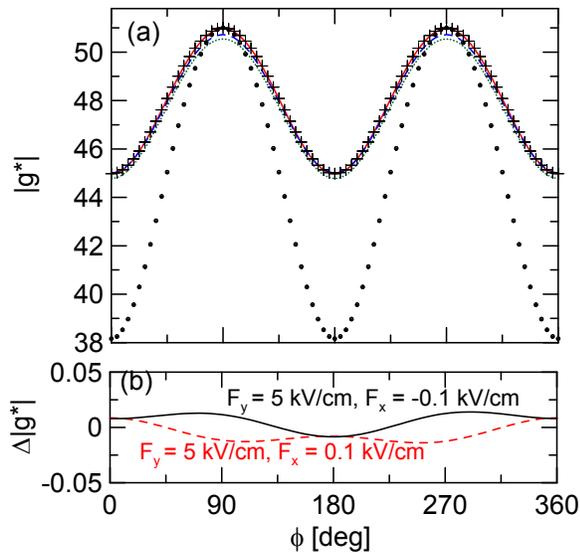}
                 \caption{(color online) (a) Effective $g$-factor obtained for a nanowire quantum dot with $R=10$ nm (red solid curve), $R=50$ nm (blue dashed curve) and $R=100$ nm (green dotted curve) obtained for $B=100$ mT. The symbols presents results obtained from Eq. (\ref{ang}) for the two lowest energy states $N$=1 (black crosses) and for the third and fourth excited states $N$=2 (black circles). (b) Difference between $g$-factor calculated for $F_x=F_y=0$, $F_z=50$ kV/cm and calculated in the presence of the electric fields in the $x$ and $y$ directions as marked on the plot.}
 \label{fig3}
\end{figure}

As the magnetic field is rotated between the easy and hard axes the spin-polarization of the states changes which results in changes of the spin-splitting strength. The latter term in Eq. (\ref{anel}) introduces Zeeman energy splitting between the energy levels of the two states. We can see that
\begin{equation}
g^* = g\sqrt{1-(1-\lambda^2+\kappa^2)\cos^2\phi}
\label{ang}
\end{equation}
 is an effective $g$-factor that is dependent on the orientation of magnetic field with the angle $\phi$. With the crosses in Fig. \ref{fig3} we plot effective $g$-factor as obtained from Eq. (\ref{ang}) along with the values obtained in the three-dimensional calculation (calculated as $g^*=\Delta E/\mu_B B$, where $\Delta E$ is the energy difference between the energy of first excited state and the ground state -- see Fig. \ref{fig0}) for different nanowire radii. For the nanowire radius $R=10$ nm the analytical solution and the result of the three-dimensional calculation match. For larger values of $R$ the shape of the dependencies comply, only the amplitude is different, with the biggest discrepancy for the wide nanowire with $R=100$ nm. The effective $g$-factor dependence obtained from the two excited states as calculated from Eq. (\ref{ang}) is plotted in Fig. \ref{fig3}(a) with circles. We observe that due to increased value of $\lambda_2$ the amplitude of the oscillation is greatly increased.

\subsection{Additional SO terms}
Additional external electric fields in the device that results from e.q. source and drain voltage difference or from gating of the nanowire can activate additional terms of the Rashba Hamiltonian, which takes the general form
\begin{equation}
\begin{split}
H_{SO}&=\alpha_0\left[F_x(\sigma_yk_z-\sigma_zk_y)\right.\\
&\left.+F_y(\sigma_zk_x-\sigma_xk_z)+F_z(\sigma_xk_y-\sigma_yk_x)\right].
\end{split}
\end{equation}
We inspect the influence of these additional terms on the anisotropic $g$-factor including in addition to $F_z = 50$ kV/cm the electric field in the $x$-direction (resulting from the bias voltage) and assuming the electric field in the $y$-direction $F_y = 5$ kV/cm. Figure \ref{fig3}(b) presents difference between results obtained with additional fields $F_x, F_y$ and results obtained for only $F_z$ present. Only slight differences are observed with the highest magnitude at the easy axes, i.e., $\phi=90^o$ and $\phi=270^o$.
\subsection{Two-electron results}

\begin{figure}[ht!]
\epsfxsize=80mm
                \epsfbox[41 51 590 790] {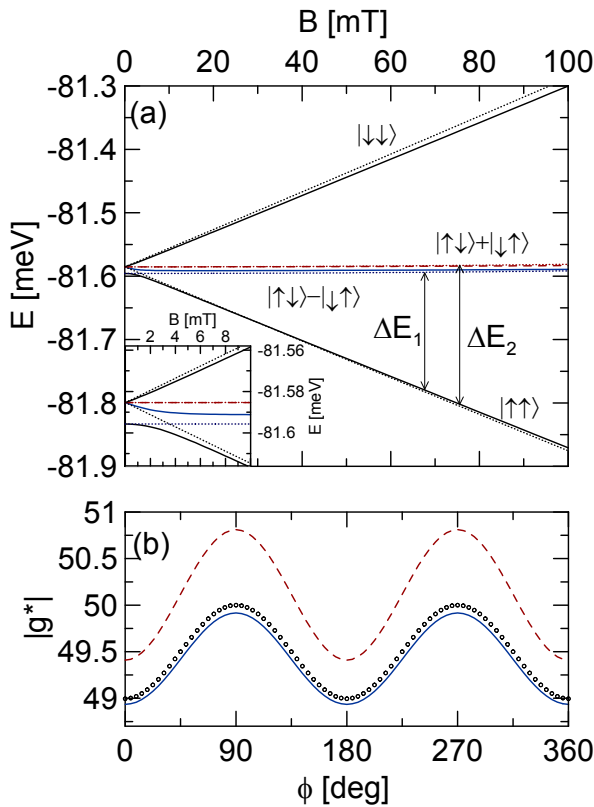}
                 \caption{(color online) (a) Two-electron energy spectrum of a coupled nanowire quantum dots with radius $R=30$ nm. Solid curves present results for $\phi=0$ and dotted curves for $\phi=90^o$.  With $|\uparrow\uparrow\rangle$, $|\downarrow\downarrow\rangle$, $|\downarrow\uparrow\rangle$, $|\uparrow\downarrow\rangle-|\downarrow\uparrow\rangle$ and $|\uparrow\downarrow\rangle+|\downarrow\uparrow\rangle$ we mark the spin configuration of the states parallel or antiparallel to the magnetic field as found without SO coupling. The inset presents the energy levels for low values of the magnetic field where the avoided crossing appears. (b) With the curves -- effective $g$-factor calculated from the energy splittings between the ground-state energy level and the energy levels depicted with blue solid ($\Delta E_1$) and red dashed ($\Delta E_2$) curves in (a) for $B=200$ mT. The circles correspond to the effective one-electron $g$-factor as obtained from Eq. \ref{ang} (shifted down by 1) for a single quantum dot with the length $L=120$ nm.}
 \label{fig4}
\end{figure}

The experimentally probed anisotropy of the $g$-factor is extracted from the slopes of resonance lines in EDSR experiments on double quantum dots in the two-electron regime.\cite{schroer,nadj2012}
Figure \ref{fig4}(a) presents two-electron energy spectrum of weakly coupled quantum dots defined in a nanowire with radius $R=30$ nm obtained in the three-dimensional calculation. Results for the magnetic field oriented along the nanowire axis with $\phi=0$ (perpendicular to the nanowire with $\phi=90^o$) are plotted with solid (dotted) curves. The confinement potential includes now a potential barrier of 60 nm width that separates the electrons in adjacent dots both of $120$ nm width. At $B=0$ the ground-state is a singlet state ($|\uparrow\downarrow\rangle-|\downarrow\uparrow\rangle$) split from the degenerate triplet states [see the inset to Fig. \ref{fig4}(a)]. We tune the barrier height to 5 meV to match the singlet-triplet separation of $\simeq 5\;\mu$eV as measured in Ref. \onlinecite{nadj2012}.

At $B=3$ mT an avoided crossing between the two lowest energy levels appears for $\phi=0$ due to spin mixing by the SO interaction. The width of the anticrossing is $\Delta E\simeq 8.2\;\mu$eV which is similar to the value measured in Ref. \onlinecite{nadj2012}, i.e., $\simeq 5\;\mu$eV. The experiment [\onlinecite{nadj2012}] established that the anticrossing vanished for $\phi=90^o$ and $\phi=270^o$ which is also the case in the present results -- the anticrossing vanishes when the magnetic field orientation is parallel to the easy axes of the spin polarization.

After the anticrossing the magnetic field splits the energy levels of the two spin-polarized triplet states ($|\uparrow\uparrow\rangle$ and $|\downarrow\downarrow\rangle$) by the Zeeman energy. The blue solid and red dashed curves in Fig. \ref{fig4}(a) whose energy does not change (after the anticrossing) with $B$ are the singlet ($|\uparrow\downarrow\rangle-|\downarrow\uparrow\rangle$) and triplet ($|\uparrow\downarrow\rangle+|\downarrow\uparrow\rangle$) states with zero spin-component in the direction along the magnetic field. Those levels are split by exchange interation\cite{nowak2012} (additional splitting of those two energy levels occurs  when the $g$-factor along the structure is not constant\cite{nadj2012,schroer,frolov}).

The magnetic field orientation (angle $\phi$): i) influences the strength of the spin polarization of the triplet states $|\uparrow\uparrow\rangle$, $|\downarrow\downarrow\rangle$ which results in a change of the slope of the corresponding energy levels and ii) change in the exchange energy (spacing between energy levels of $|\uparrow\downarrow\rangle-|\downarrow\uparrow\rangle$ and $|\uparrow\downarrow\rangle+|\downarrow\uparrow\rangle$ states, plotted with blue solid and red dashed curves in Fig. \ref{fig4}(a). These two effects lead to a dependence of the effective $g$-factor on $\phi$ which we calculate from the energy splittings between the ground-state and the first and second excited states and plot in Fig. \ref{fig4}(b) with the blue solid and red dashed curves respectively. We find that the shape of both curves in Fig. \ref{fig4}(b) match the shape of the single-electron dependence presented in Fig. \ref{fig3} only the amplitude of the oscillations is lower. As described by Eq. (\ref{ang}) for the single-electron case the amplitude of $g$-factor oscillations depends on the dot length. In the present case each of the coupled quantum dots has a length of $L=120$ nm.  The effective $g$-factor obtained for a {\it single} dot of this length as calculated from Eq. (\ref{ang}) is plotted with circles in Fig. \ref{fig4}(b). Obtained oscillations have similar amplitude to the ones obtained for the two-electron system. This suggest that the low amplitude in the two-electron case results from the fact that each electron resides in a separate dot and the shape of the oscillations is controlled mainly by the single-electron spin polarization anisotropy process described previously.

The shape of the $g$-factor dependence is similar to the one obtained in the experiment [\onlinecite{nadj2012}]. In particular an agreement is obtained in the context of the slight change of the oscillation amplitude of the red dashed and blue solid curves in Fig. \ref{fig4}(b). This difference in amplitudes is due to modification of the exchange energy that separates the energy levels of the singlet ($|\uparrow\downarrow\rangle-|\downarrow\uparrow\rangle$) and triplet ($|\uparrow\downarrow\rangle+|\downarrow\uparrow\rangle$) states by the rotated magnetic field. However, the experimental dependence of the effective $g$-factor is shifted (with minima at $\phi=124^o$ and $\phi=304^o$) with respect to the present result. We performed calculations for quantum dots in a nanowire of larger radius ($R=100$ nm) ruling out the possible orbital effects of the magnetic field as a reason for the shift. Also the additional terms of Rashba coupling are not responsible for such a shift as discussed in the Subsection C. On the other hand the $g$-factor in quantum dots is affected by the local strain and asymmetries in the structure\cite{pryor} which can influence the $g$-factor as an concurrent process to the anisotropic spin polarization.


\section{Summary and conclusions}

In the present work we studied the anisotropy of the spin polarization in a narrow nanowire quantum dot in the presence of SO coupling. Solving three-dimensional Schr\"odinger equation we showed that the strength of the spin polarization in the presence of Rashba SO interaction depends on the orientation of the magnetic field and there are hard and easy spin polarization axes. We explained the existence of these axes by the intrinsic tendency of SO coupling to polarize spins in the direction perpendicular to the nanowire. For the magnetic field aligned in this direction the electron spin polarization can be nearly complete depending on the nanowire radius. We presented analytical solution for the one-dimensional limit where the spin polarization can be complete and compared its results with the calculation for a finite thickness nanowire. The spin polarization anisotropy results in the effective $g$-factor dependence on the magnetic field orientation which is stronger for the excited states. The anisotropy of the single electron spin polarization results in the changes of avoided crossing width in the lowest part of the two-electron energy spectra. The magnitude and position of extrema of this dependence matches the ones founds in the experiment. Also the form of the $g$-factor dependence resembles the one obtained in the experimental studies.

\section*{Acknowledgements}
This work was supported by the funds of Ministry of Science an Higher Education (MNiSW) for 2012 -- 2013 under Project No. IP2011038671, and by PL-Grid Infrastructure.
M.P.N. is supported by the Foundation for Polish Science (FNP) scholarship under START and the MPD Programme co-financed by the EU European Regional Development Fund.


\begin{thebibliography}{00}
\bibitem{enote} The obtained energy shift is two-times smaller that the one obtained for two-dimensional systems in Ref. [\onlinecite{serra}].
\bibitem{comment1} In a general the form of $\lambda_N$ and $\kappa_N$ can be provided for any confinement potential for which the analytical form of the wave functions are known, in particular for the harmonic oscillator potential.
\end{thebibliography}
\end{document}